\documentclass[aps,prc,preprint,amsmath,amssymb,showpacs,preprintnumbers,superscriptaddress]{revtex4-1}
\usepackage{CJK}
\usepackage{graphicx}
\usepackage{dcolumn}
\usepackage{bm}
\usepackage{color}
\usepackage{hyperref}

\allowdisplaybreaks

\begin{document}


\title{ Microscopic model for yields and total kinetic energy in nuclear fission}

\author{B. Li }
\affiliation{State Key Laboratory of Nuclear Physics and Technology, School of Physics, Peking University, Beijing 100871, China}
\author{D. Vretenar}
\email{vretenar@phy.hr}
\affiliation{Physics Department, Faculty of Science, University of Zagreb, 10000 Zagreb, Croatia}
\affiliation{State Key Laboratory of Nuclear Physics and Technology, School of Physics, Peking University, Beijing 100871, China}
\author{T. Nik\v si\' c}
\affiliation{Physics Department, Faculty of Science, University of Zagreb, 10000 Zagreb, Croatia}
\affiliation{State Key Laboratory of Nuclear Physics and Technology, School of Physics, Peking University, Beijing 100871, China}
\author{P. W. Zhao }
\email{pwzhao@pku.edu.cn}
\affiliation{State Key Laboratory of Nuclear Physics and Technology, School of Physics, Peking University, Beijing 100871, China}
\author{J. Meng }
\email{mengj@pku.edu.cn}
\affiliation{State Key Laboratory of Nuclear Physics and Technology, School of Physics, Peking University, Beijing 100871, China}

\date{\today}

\begin{abstract}

An extension of time-dependent density functional theory (TDDFT), the generalized time-dependent generator coordinate method (TDGCM), is applied to a study of induced nuclear fission dynamics. In the generalized TDGCM, the correlated nuclear wave function is represented as a coherent superposition of time-dependent DFT trajectories. In the first realistic application, a large basis of 25 TDDFT trajectories is employed to calculate the charge yields and total kinetic energy distribution for the fission of $^{240}$Pu. The results are compared with available data, and with those obtained using a standard TDDFT, that does not consider quantum fluctuations, and the adiabatic TDGCM+GOA (Gaussian overlap approximation). It is shown that fragment yields and kinetic energies can simultaneously be described in a consistent microscopic framework that includes fluctuations in the collective degrees of freedom and the one-body dissipation mechanism.
\end{abstract}

\maketitle


Time-dependent density functional theory (TDDFT) \cite{Rung1984RPL,Ullrich2011} has been very successfully applied to studies of spectroscopy and dynamics in chemistry, biology, physics, and material sciences. In nuclear physics, for instance, several  implementations of TDDFT have been employed in the description of small and large amplitude collective motion. A characteristic example of the latter is nuclear fission \cite{Schunck16PPNP,Schunck22PPNP,Bender2020JPG}.
Various aspects of the fission process have been analyzed in the TDDFT framework
\cite{simenel12,simenel18,nakatsukasa16,stevenson19,bulgac16,magierski17,scamps18,Stetcu2021PRL,Bulgac2022PRL,Ren_22PRL,Ren2022b,Ren_22PRC,Li2023PRC_FTTDDFT,Scamps2023PRC}.
Even though nuclear TDDFT incorporates the one-body dissipation mechanism, it can only model a single fission event at a time, by propagating independent nucleons in self-consistent mean-field potentials. This is because at each time the nuclear wave function is represented by a single, time-dependent many-body product state. Therefore, in its standard formulation,TDDFT does not include  quantum fluctuations of collective degrees of freedom. To develop fully microscopic methods that provide accurate predictions of fission observables, such as charge and mass yields, total kinetic energy, and angular momentum of fragments, it is necessary to incorporate quantum fluctuations in a dynamical modeling of the fission precess. Approximate schemes to include quantum fluctuations in TDDFT-based models have been considered \cite{Tanimura2017PRL,qiang21}, based on the introduction of phenomenological stochastic terms.

Quantum fluctuations of collective degrees of freedom can naturally be included in the time-dependent generator coordinate method (TDGCM), by considering the evolution of a set of characteristic coordinates in collective space~\cite{krappe12,younes19,Regnier2016_PRC93-054611,Tao2017PRC,Verriere2020_FP8-233}. In TDGCM
the nuclear wave function is represented by a superposition of generator states that are functions of collective coordinates, and can be applied to an adiabatic description of the entire fission process, from the quasi-stationary initial state to scission. This method is fully quantum mechanical but only takes into account collective degrees of freedom and, thus, cannot be used to describe the highly dissipative dynamics, that is, energy dissipation from collective to nucleonic degrees of freedom, that occurs beyond the saddle point \cite{Tanimura2015PRC,Tanimura2017PRL}. Several attempts to include a dissipation mechanism in TDGCM have been reported~\cite{Dietrich2010NPA,Zhao2022PRC,Zhao2022PRC_TKE}, but the corresponding models are
difficult to implement and often computationally prohibitive for realistic calculations of fission properties. An important step toward a rigorous and quantitative TDGCM description of fission dynamics, based on the Schr\"odinger collective intrinsic model (SCIM) \cite{Bernard2011PRC} that goes beyond the adiabatic approximation, has recently been reported in Ref.~\cite{Carpentier2024PRL}, where a class of methods is developed to construct continuous potential energy surfaces, both adiabatic and including excited states, of many-body quantum systems.

TDDFT and TDGCM, therefore, present complementary frameworks for the description of fission dynamics. This has motivated attempts to develop a unified method that extends the standard TDDFT, simultaneously including quantum fluctuations and  dissipation. Very recently, the generalized TDGCM has been implemented and applied to the dynamics of small amplitude collective motion of atomic nuclei \cite{Li2023PRC_gdTDGCM,Marevicc2023PRC}. In this approach, the nuclear wave function is expressed as a superposition of many generator states, and both the generator states and their weight functions explicitly depend on time. We have also used this model in an exploratory study of induced fission of $^{240}$Pu, with the inclusion of pairing correlations and quantum superposition effects among sets of TDDFT generating fission trajectories~\cite{Li2024FOP}.

In this letter, nuclear fission dynamics is analyzed quantitatively by using a newly developed generalized TDGCM model,
and the total kinetic energy and charge yields of fragments are calculated without any adjustable parameters beyond those that determine the energy density functional and pairing interaction.
Taking $^{240}$Pu as an example, the nuclear wave function is expressed as a superposition of a relatively large number of  time-dependent DFT product states, and evolved in time by the generalized TDGCM, taking into account
effects of quantum fluctuations and the dissipative mechanism that couples collective and single-nucleon degrees of freedom.
For the details of the generalized TDGCM, we refer the reader to the Supplement~\cite{Li2024Supp} and Refs~\cite{Li2023PRC_gdTDGCM,Li2024FOP}.

In generalized TDGCM, the nuclear wave function reads ~\cite{Regnier2019PRC,Li2023PRC_gdTDGCM,Li2024Supp,Marevicc2023PRC,Li2024FOP}
\begin{equation}
    |\Psi(t)\rangle=\sum_{\bm q} f_{\bm q}(t) |\Phi_{\bm q}(t)\rangle,
    \label{Eq_collec_wfs}
\end{equation}
where the vector ${\bm q}$ denotes discretized {\em generator coordinates} that parametrize the collective degrees of freedom.
The time evolution of the {\em generator state} $|\Phi_{\bm q}\rangle$ is determined by time-dependent covariant density functional theory~\cite{Ren2020PLB,Ren2020PRC,Ren_22PRL,Ren_22PRC}, using the time-dependent BCS approximation for pairing correlations~\cite{ebata10TDBCS,Scamps13TDBCS}, and the weight functions satisfy the time-dependent Hill-Wheeler equation~\cite{Regnier2019PRC},
\begin{equation}
    i\hbar \mathcal{N}\partial_t f=(\mathcal{H}-\mathcal{H}^{MF}) f \;.
    \label{TD-HW-f}
\end{equation}
The time-dependent kernels $\mathcal{N}$, $\mathcal{H}$, and $\mathcal{H}^{MF}$ include the overlap, the Hamiltonian, and the time derivative of the generator states, respectively:
\begin{subequations}
 \begin{align}
&\mathcal{N}_{\bm{q'q}}(t)=\langle\Phi_{\bm {q'}}(t)|\Phi_{\bm q}(t)\rangle,\label{Eq_N}\\
&\mathcal{H}_{\bm{q'q}}(t)=\langle\Phi_{\bm {q'}}(t)|\hat{H}|\Phi_{\bm q}(t)\rangle,\label{Eq_H}\\
&\mathcal{H}^{MF}_{\bm{q'q}}(t)=\langle\Phi_{\bm {q'}}(t)|i\hbar\partial_t|\Phi_{\bm q}(t)\rangle.
\label{Eq_H_mf}
 \end{align}
\end{subequations}
In the present analysis, for instance, the kernels $\mathcal{N}$, $\mathcal{H}$, and $\mathcal{H}^{MF}$ are $25 \times 25$ time-dependent matrices that are computed at each step of the time evolution.

The weight function $f_{\bm q}$ does not represent the probability amplitude of finding the system at the collective coordinate ${\bm q}$. The corresponding collective wave function $g_{\bm q}$ is defined by the transformation $ g=\mathcal{N}^{1/2}f$~\cite{Reinhard1987RPP}, and governed by the time-dependent equation
\begin{equation}
    i\hbar \dot{g}=\mathcal{N}^{-1/2}(\mathcal{H}-\mathcal{H}^{MF})\mathcal{N}^{-1/2}g+i\hbar\dot{\mathcal{N}}^{1/2}\mathcal{N}^{-1/2}g \;,
    \label{TD-HW-g}
\end{equation}
where $\mathcal{N}^{1/2}$ is the square root of the overlap kernel matrix. For the correlated nuclear wave function, the expectation value of an observable $\hat{O}$ reads
\begin{equation}
\langle \Psi(t)|\hat{O}|\Psi(t)\rangle =\sum_{{\bm q}{\bm q\prime}}f_{\bm q\prime}^{*}(t)f_{\bm q}(t) \langle\Phi_{\bm q\prime}(t) |\hat{O} |\Phi_{\bm q}(t)\rangle.
\label{Eq_observable}
\end{equation}

\begin{figure}[!htbp]
\centering
\includegraphics[width=0.8\textwidth]{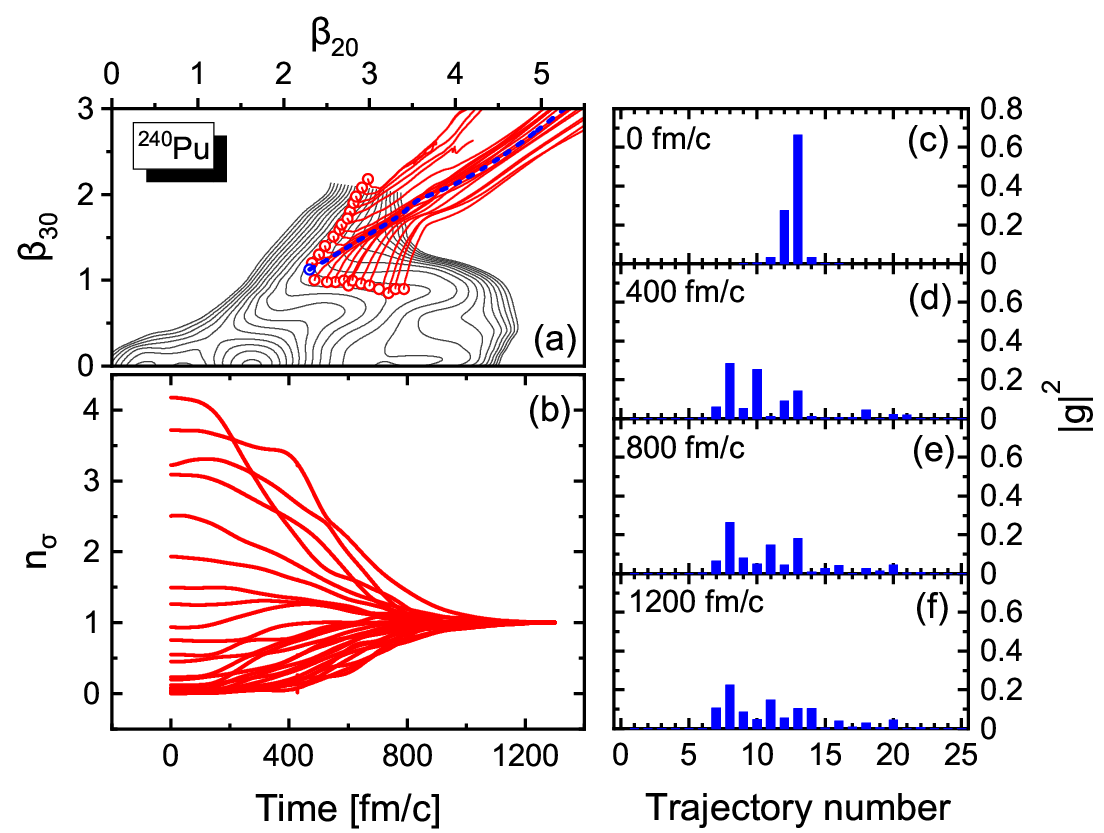}
\caption{(color online). Panel~(a): Self-consistent deformation energy surface of $^{240}$Pu in the plane of quadrupole-octupole axially-symmetric deformation parameters, calculated with the relativistic density functional PC-PK1
~\cite{zhao10} and a monopole pairing interaction. Contours join points on the surface with the same energy, and the open dots correspond to points on the iso-energy curve at 1 MeV below the energy of equilibrium minimum. The curves correspond to self-consistent TDDFT fission trajectories that start from the 25 initial points, and are used as a time-dependent generator basis for the generalized TDGCM. Starting from the largest value of the octupole moment, the initial points are labelled from 1 to 25. The dashed curve denotes trajectory number 13, and its initial point is at $(\beta_{20},\beta_{30})=(2.30,1.13)$.~Panel~(b): Time evolution of the eigenvalues of the norm kernel. Panel~(c)$-$(f): Square moduli of the components of the TDGCM collective wave function, that starts from the initial point $(\beta_{20},\beta_{30})=(2.30,1.13)$ of trajectory number 13, at $0$, $400$, $800$, and $1200~{\rm fm}/c$.}
\label{fig_ES}
\end{figure}

In panel (a) of Fig.~\ref{fig_ES}, we display the self-consistent deformation energy surface of $^{240}$Pu as function of the axial quadrupole ($\beta_{20}$) and octupole ($\beta_{30}$) deformation parameters,
which is obtained by self-consistent deformation-constrained relativistic DFT calculations in a three dimensional lattice space~\cite{Ren2017PRC,Ren2017PRC, ren19LCS, ren20_NPA,Li2020PRC,Xu2024PRC,Xu2024PLB}.
The trajectories of 25 time-dependent generator states,
which start at the initial points denoted by open circles, are subsequently evolved by TDDFT. Since it effectively describes the classical evolution of independent nucleons in self-consistent mean-field potentials, this method cannot be
applied in the classically forbidden region of the collective space. The starting points for the TDDFT evolution are usually taken beyond the outer barrier, and here they are located along an iso-energy curve 1 MeV below the energy of the equilibrium minimum. This choice ensures that most trajectories lead to scission, even without boosting the initial wave functions. We use the labels $1-25$ for the time-dependent generator states and their initial points, starting from the largest initial octupole deformation $\beta_{30}$.
As a characteristic example, the dashed curve corresponds to trajectory number 13 which starts from the initial point $(\beta_{20},\beta_{30})=(2.31,1.13)$, and is propagated in time by TDDFT with the functional PC-PK1~\cite{zhao10} and a monopole pairing interaction. When the nucleus eventually scissions along this trajectory, average properties of the two fragments, such as charge number, mass number, and TKE, can be computed but, obviously, fluctuations in the collective coordinates are not taken into account.
Starting from the same initial states, the fission process can also be modeled by the generalized TDGCM. In this framework the collective wave function is, at all times, a coherent superposition of the 25 TDDFT trajectories, and it is evolved by Eq.~(\ref{TD-HW-g}),
\begin{equation}
g_{\bm q}(t) =\sum_{{\bm q\prime =1}}^{25} \mathcal{N}^{1/2}_{{\bm q}{\bm q\prime}}(t) f_{\bm q\prime}(t).
\end{equation}
The eigenvalues of the overlap kernel matrix, as functions of time, are shown in the panel~(b) of Fig.~\ref{fig_ES}. One notices that these eigenvalues gradually
approach 1 with time, which means that the TDDFT trajectories become orthogonal.
This is because TDDFT trajectories are independent of other trajectories and correspond to distinct pairs of fragments with different particle numbers, and at different locations after scission.

In panels (c)$-$(f) we show the square moduli of the components of the TDGCM collective wave function, that starts from the initial point $(\beta_{20},\beta_{30})=(2.30,1.13)$ of trajectory number 13, at $0$, $400$, $800$, and $1200~{\rm fm}/c$, respectively.
These square moduli $|g(q)|^2,~q=1,2,...,25$, where $q$ is the trajectory number, correspond to the probability of the $q$-th TDDFT trajectory.
At the initial time $t = 0~{\rm fm}/c$, the components of the collective wave function are concentrated in the vicinity of trajectory number 13, and then spread out during the time evolution. Note that in TDDFT the probability of a trajectory is either $1$ or $0$,  because the nuclear wave function is only represented by a single product state.

The dispersion of the collective wave function is most pronounced during the initial interval, $0-400~{\rm fm}/c$, and this indicates that quantum fluctuation effects are important well before scission.
There are only small changes in the collective wave function after $800~{\rm fm}/c$, because the time-dependent generator states start to become orthogonal. This analysis can be extended to the collective wave functions that start from the other initial states.
The corresponding square moduli of the TDDFT components of 25 collective wave functions $|g|^2$ at $1300~{\rm fm}/c$, when the fragments are completely separated for most TDDFT trajectories, are displayed in panels~(1)$-$(25) of Fig.~\ref{fig_g}, respectively.
The bars, normalized to $1$, denote the components of the collective wave functions obtained from the generalized TDGCM trajectories that start at the same initial points as the TDDFT trajectories,
but represent a coherent superposition of all 25 TDDFT trajectories.

\begin{figure}[!htbp]
\centering
\includegraphics[width=1\textwidth]{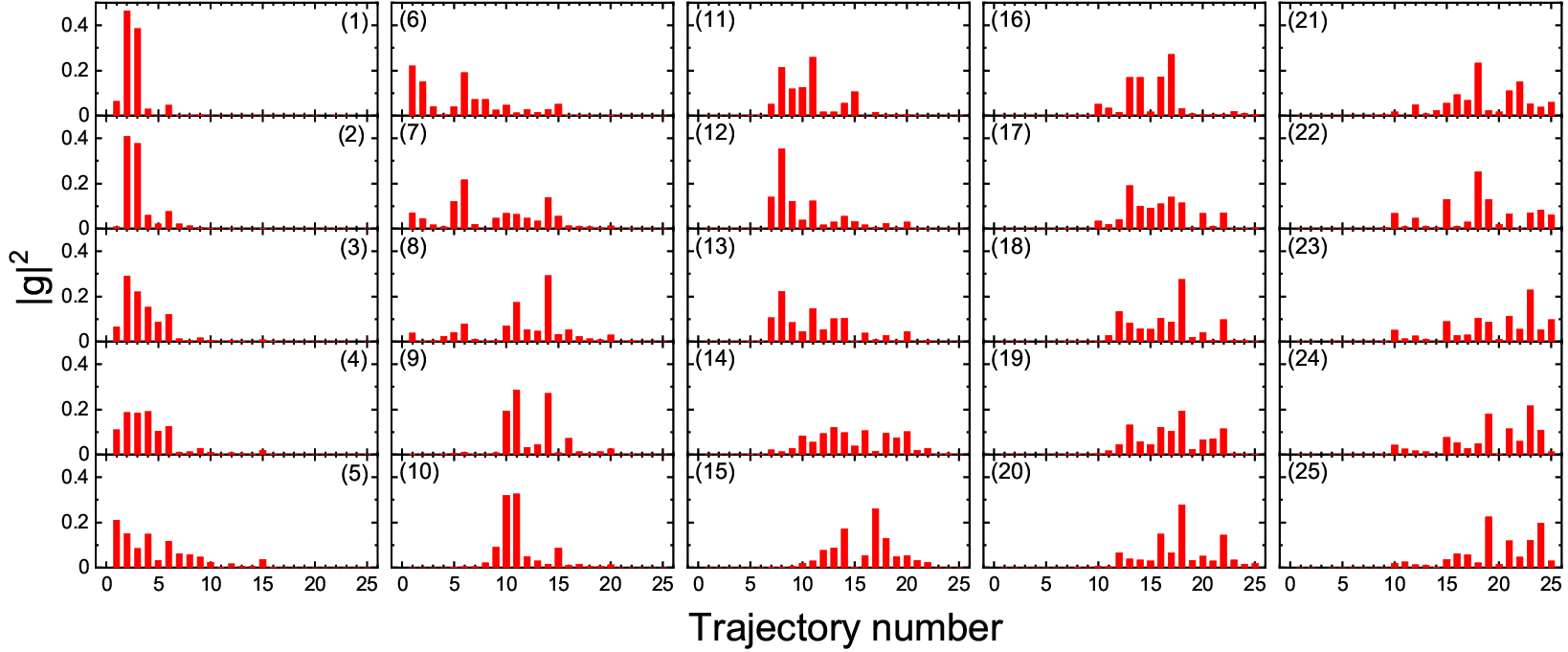}
\caption{(color online) The square moduli of the 25 TDDFT components of the generalized TDGCM collective wave functions $|g|^2$, at time $1300~{\rm fm}/c$. The generalized TDGCM trajectories $1-25$ start from the initial points $1-25$, shown in panel~(a) of Fig~\ref{fig_ES}.}
\label{fig_g}
\end{figure}

To obtain the charge yields for the correlated nuclear wave function $|\Psi\rangle$ at $t=1300~{\rm fm}/c$, which is a superposition of 25 TDDFT generator states and includes paring correlations, we employ particle number projection \cite{Scamps2013PRC}.
The probability of finding $z$ protons in the subspace $V_f$ that corresponds to one of the fragments, when the total system contains $Z$ protons, reads
\begin{equation}
   P(z|Z,t)=\frac{\langle\Psi(t)\left|\hat{P}_z^{V_f}\hat{P}_Z\right|\Psi(t) \rangle}{\langle\Psi(t) \left|\hat{P}_Z\right|\Psi (t)\rangle}=
   \frac{\sum_{{\bm q}{\bm q\prime}}f_{\bm q\prime}^{*}(t)f_{\bm q}(t) \langle\Phi_{\bm q\prime}(t)\left|\hat{P}_z^{V_f}\hat{P}_Z\right|\Phi_{\bm q}(t)\rangle}%
   {\sum_{{\bm q}{\bm q\prime}}f_{\bm q\prime}^{*}(t) f_{\bm q}(t) \langle\Phi_{\bm q\prime}(t) \left|\hat{P}_Z \right|\Phi_{\bm q}(t)\rangle},%
   \label{PNP}
\end{equation}
where $\hat{P}_z^{V_f}$ ($\hat{P}_Z$) is the projection operator on a given number of protons $z$ ($Z$) inside the subspace $V_f$ (entire space). This expression is, of course, valid also for the number of neutrons, and we refer the reader to the Supplement \cite{Li2024Supp} for detailed formulas.
The proton probability distributions for the wave functions that start from the initial points $1-25$, and are evolved by the generalized TDGCM to $t=1300~{\rm fm}/c$,
are shown in panels (1)$-$(25) of Fig.~\ref{fig_p}, respectively. They are normalized to 1 for the light and heavy fragments.
As an example, let us consider the initial state with the largest value of $\beta_{30}$, i.e., the first initial state. It mainly contributes to fragments with the proton number $Z=36,37,38$ and $56,57,58$,
with significant contributions for $Z=42,43$ and $Z=51,52$.
We note that the corresponding TDDFT state that starts from the same point, and is evolved by TDDFT, does not lead to scission until $2000~{\rm fm}/c$.
Decreasing the initial $\beta_{30}$, the proton probability distributions of generalized TDGCM trajectories gradually concentrate in the region $Z=40,41,42$ and $52,53,54$. Finally, to obtain the total charge yields of $^{240}$Pu and compare with data, we sum and normalize the proton distributions from all 25 TDGCM trajectories.

\begin{figure}[!htbp]
\centering
\includegraphics[width=1\textwidth]{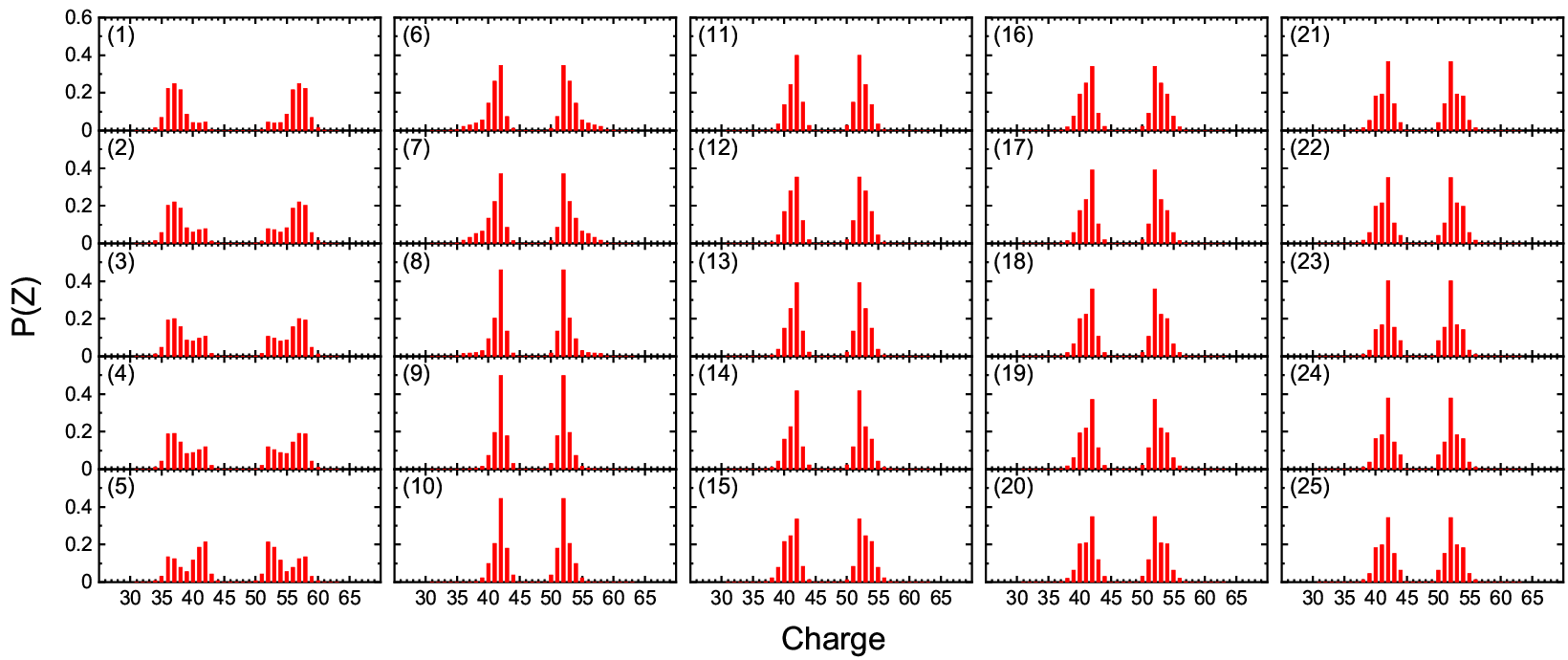}
\caption{(color online). Same as Fig.~\ref{fig_g}, but for probability distributions of proton number. }
\label{fig_p}
\end{figure}

\begin{figure}[!htbp]
\centering
\includegraphics[width=0.55\textwidth]{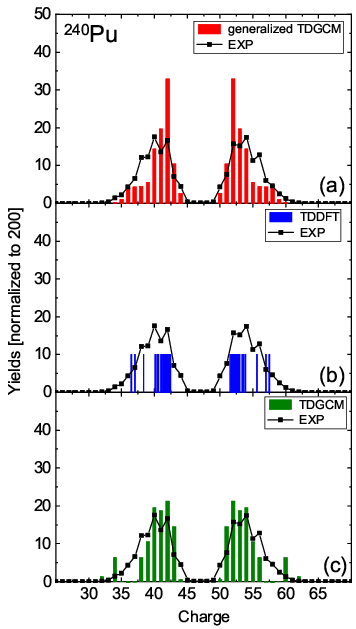}
\caption{(Color online). Charge yields for induced fission of $^{240}$Pu. The yields computed with the generalized TDGCM (a), TDDFT (b), and TDGCM + GOA (c)~\cite{Ren_22PRC},
 are shown in comparison with the experimental charge distribution. The data are from Ref.~\cite{Ramos2018_PRC97-054612}, and correspond to an average excitation energy of 10.7 MeV.}
\label{fig_CY}
\end{figure}

The charge yields of $^{240}$Pu obtained by the generalized TDGCM with particle number projection are shown in panel (a) of Fig.~\ref{fig_CY}. The experimental result is reproduced without any parameter adjustment. Because the nuclear wave function in TDDFT is only represented by a single product state, each TDDFT trajectory produces a pair of fragments, shown as
as bars in panel (b) of Fig.~\ref{fig_CY}.
TDDFT trajectories predominantly lead to fragments with charge numbers around $Z=41$ and $53$, in general agreement with data. Finally, in panel (c) of Fig.~\ref{fig_CY} we display the yields predicted by the standard TDGCM plus Gaussian overlap approximation (TDGCM+GOA) \cite{Ren_22PRC}. Compared to the results obtained with the generalized TDGCM in panel (a), it appears that TDGCM+GOA does not reproduce so well the data in the tails of the distribution, and also for more symmetric fission events. This result, of course, corresponds to the specific example considered here. In a more systematic study, yields predicted by the generalized TDGCM and TDGCM+GOA should be compared for a series of fissioning nuclides.

The total kinetic energies (TKE), computed by the generalized TDGCM, TDDFT, and TDGCM+GOA are shown in Fig.~\ref{fig_TKE}. For a single TD-DFT trajectory, the total kinetic energy (TKE) at a finite distance between the fission fragments ($\approx 25$ fm, at which shape relaxation brings the fragments to their equilibrium shapes) is calculated using the expression
\begin{equation}
E_{TKE}=\frac{1}{2}mA_{\rm H}^{\bm q}\bm{v}^2_{{\rm H},{\bm q}}+\frac{1}{2}mA_{\rm L}^{\bm q}\bm{v}^2_{{\rm L},{\bm q}}+E_{\rm Coul}^{\bm q},
\end{equation}
where the velocity of the fragment $f=H,L$ reads
\begin{equation}
\bm{v}_{f,{\bm q}}=\frac{1}{mA_{f}^{\bm q}}~\int_{V^{\bm q}_f} d\bm{r}~\bm{j}_{\bm q}(\bm{r}) \;,
\end{equation}
and $\bm{j}(\bm{r})$ is the total current density. The integration is over the half-volume corresponding to the fragment $f$, and $E_{\mathrm{Coul}}$ is the Coulomb energy. For each correlated generalized TD-GCM trajectory,
the average charge number of the fragments and the total kinetic energy are calculated using Eq.~(\ref{Eq_observable}) for the expectation value of the corresponding observable. In the case of TDGCM+GOA, the kinetic energy of the fragments corresponds to just their Coulomb repulsion at scission. This is because in the adiabatic approximation, on which TDGCM+GOA is based, all the potential energy is converted into collective kinetic energy during the saddle-to-scission evolution \cite{Ren_22PRC}. The nascent fragments are cold and, as shown in Fig.~\ref{fig_TKE}, the calculated TKEs are systematically too large when compared to data.

In generalized TDGCM and TDDFT, because the one-body dissipation mechanism is automatically included, part of the collective flow energy is converted to intrinsic energy and heats up the fissioning nucleus~\cite{Bulgac2019,Li2023PRC_FTTD}, thus producing hot excited fragments. Compared with data~\cite{Caamano2015PRC}, both the generalized TDGCM and TDDFT reproduce the experimental TKEs for fragments close to the peaks of the charge yield distribution, but underestimate the TKEs for the tails of the distribution. As already noted in Ref.~\cite{Ren_22PRC}, this is partly due to the fact that the calculated TKEs do not include the contribution of pre-scission energy because the initial points for the fission trajectories are on the deformation energy surface, while the data correspond to an average excitation energy of the fissioning nucleus of 9 MeV \cite{Caamano2015PRC}. Thus, the calculated TDGCM and TDDFT values shown in Fig.~\ref{fig_TKE} present a lower bound for the total kinetic energies, and can be further improved by including the excitation energy of the nucleus at the initial points of time evolution.

\begin{figure}[!htbp]
\centering
\includegraphics[width=0.65\textwidth]{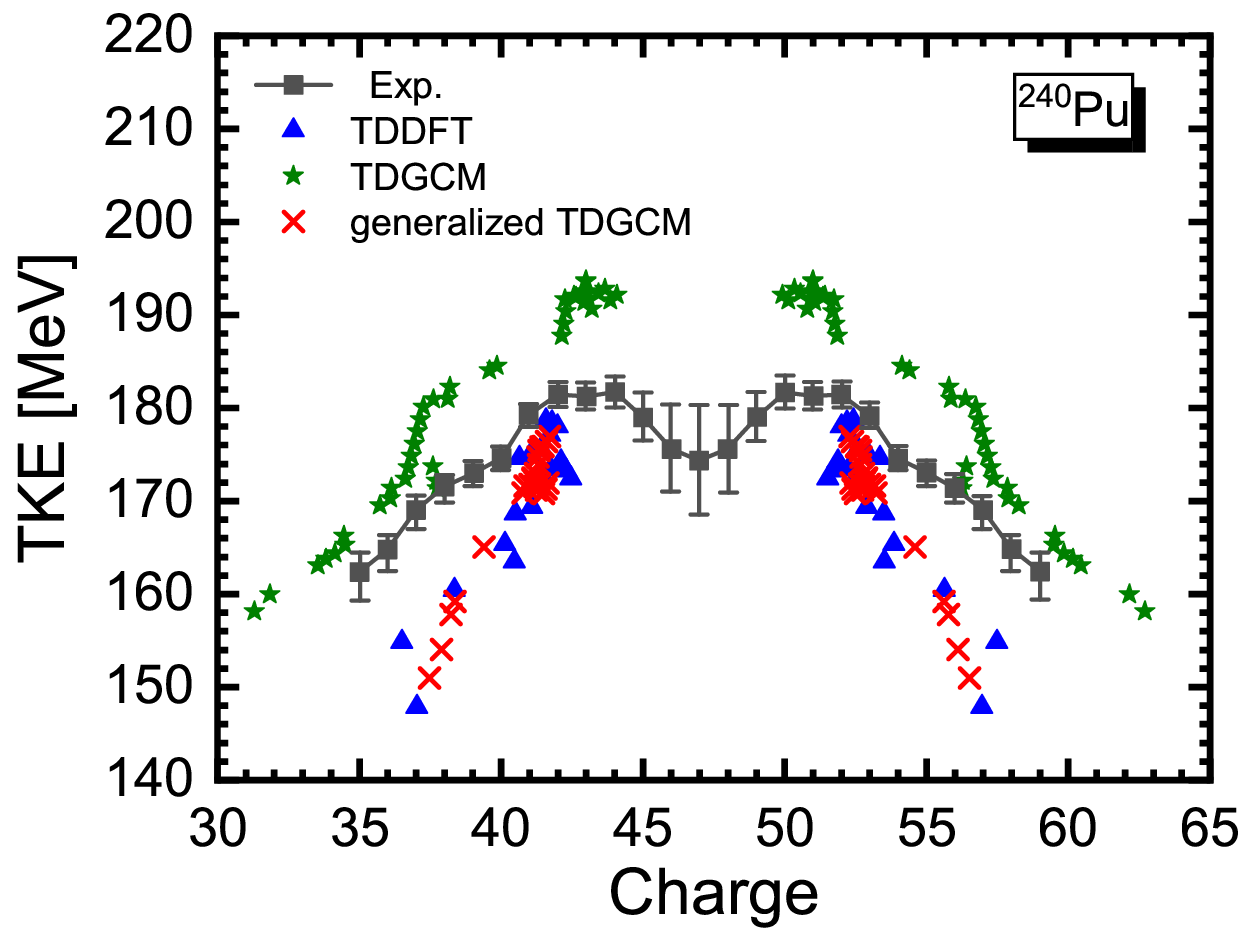}
\caption{(Color online). Total kinetic energies of the emerging fragments for induced fission of $^{240}$Pu, as functions of the fragment charge. The TDDFT, TDGCM+GOA~\cite{Ren_22PRC}, and generalized TDGCM
results are shown in comparison to the data~\cite{Caamano2015PRC}. }
\label{fig_TKE}
\end{figure}

In summary, an extension of time-dependent density functional theory, based on the time-dependent generator coordinate method, has been applied to nuclear fission dynamics. In the first realistic application to induced fission of $^{240}$Pu, a large basis of 25 TDDFT trajectories has been used to calculate the charge yields and total kinetic energy distribution. The effects of quantum fluctuations in the collective degrees of freedom and the one-body dissipation mechanism, for the first time simultaneously included in a consistent microscopic framework, have been analyzed in comparison with experimental values, and results obtained with standard TDDFT and the adiabatic TDGCM+GOA. The TDGCM-based extension of standard TDDFT, presented in this work, can be applied to other physical processes in which quantum fluctuations are essential for a correct description of dynamics.
\begin{acknowledgments}
This work was supported in part by the High-End Foreign Experts Plan of China,
the National Natural Science Foundation of China (Grants No.11935003, No.12435006, No.12475117, and No.12141501),
the State Key Laboratory of Nuclear Physics and Technology, Peking University (Grant No. NPT2023ZX03),
the National Key Laboratory of Neutron Science and Technology (Grant No. NST202401016),
the National Key R\&D Program of China 2024YFE0109803,
the High-Performance Computing Platform of Peking University,
by the ``Scientific Center of Excellence for Quantum and Complex Systems, and Representations of Lie Algebras'', PK.1.1.02, European Union, European Regional Development Fund, and by the Croatian Science Foundation under the project Relativistic Nuclear Many-Body Theory in the Multimessenger Observation Era (IP-2022-10-7773).
\end{acknowledgments}

\clearpage
\bigskip
%

\clearpage

\title{Supplemental Material for ``Microscopic model for yields and total kinetic energy in induced nuclear fission''}
\author{B. Li }
\affiliation{State Key Laboratory of Nuclear Physics and Technology, School of Physics, Peking University, Beijing 100871, China}
\author{D. Vretenar}
\email{vretenar@phy.hr}
\affiliation{Physics Department, Faculty of Science, University of Zagreb, 10000 Zagreb, Croatia}
\affiliation{State Key Laboratory of Nuclear Physics and Technology, School of Physics, Peking University, Beijing 100871, China}
\author{T. Nik\v si\' c}
\affiliation{Physics Department, Faculty of Science, University of Zagreb, 10000 Zagreb, Croatia}
\affiliation{State Key Laboratory of Nuclear Physics and Technology, School of Physics, Peking University, Beijing 100871, China}
\author{P. W. Zhao }
\email{pwzhao@pku.edu.cn}
\affiliation{State Key Laboratory of Nuclear Physics and Technology, School of Physics, Peking University, Beijing 100871, China}
\author{J. Meng }
\email{mengj@pku.edu.cn}
\affiliation{State Key Laboratory of Nuclear Physics and Technology, School of Physics, Peking University, Beijing 100871, China}

\date{\today}

\maketitle

This supplemental information contains:
\begin{itemize}
\item A description of the implementation of the generalized time-dependent generator coordinate method (TDGCM) used in the present study.
\item A description of the particle number projection method.
\end{itemize}
\section{SUPPLEMENTAL METHODS: Generalized time-dependent GCM}

The TD-GCM correlated nuclear wave function with discretized generator coordinates reads ~\cite{Regnier2019PRC,Li2023PRC_gdTDGCM,Marevicc2023PRC,Li2024FOP}
\begin{equation}
    |\Psi(t)\rangle=\sum_{\bm q} f_{\bm q}(t) |\Phi_{\bm q}(t)\rangle,
    \label{Eq_collec_wfs}
\end{equation}
where the vector ${\bm q}$ denotes {\em generator coordinates} that parametrize collective degrees of freedom.
This wave function is a linear superposition of, generally non-orthogonal, many-body {\em generator states} $|\Phi_{\bm q}(t)\rangle$, and $f_{\bm q}(t)$ are the corresponding complex-valued {\em weight functions}.
It is the solution of the time-dependent equation
\begin{equation}
    i\hbar\partial_t|\Psi(t)\rangle=\hat{H}|\Psi(t)\rangle,
    \label{Eq_td_eq}
\end{equation}
where $\hat{H}$ is the Hamiltonian of the nuclear system.

\subsection{Time evolution of generator states $|\Phi_{q}(t)\rangle$}
The evolution in time of the quasiparticle vacuum characterized by a vector of generator coordinates ${\bm q}$
\begin{equation}
   |\Phi_{\bm q}(t)\rangle = \prod_{k>0}[\mu_{{\bm q},k}(t)+\nu_{{\bm q},k}(t)c_{{\bm q},k}^\dagger(t)c_{{\bm q},\bar{k}}^\dagger(t)]|-\rangle \;,
\label{Eq_Slater}
\end{equation}
is modeled by the time-dependent covariant density functional theory~\cite{Ren_22PRL,Ren_22PRC}, using the time-dependent BCS approximation~\cite{ebata10TDBCS,Scamps13TDBCS}.
In Eq.~(\ref{Eq_Slater}), $\mu_{{\bm q},k}(t)$ and $\nu_{{\bm q},k}(t)$ are the parameters of the transformation between the canonical and quasiparticle bases, and $c_{{\bm q},k}^\dagger(t)$ denotes the creation operator associated with the canonical state $\phi_k^{\bm q}(\bm{r},t)$.~The time evolution of $\phi_k^{\bm q}(\bm{r},t)$ is determined by the time-dependent Dirac equation
\begin{equation}\label{Eq_td_Dirac_eq_BCS}
  i\frac{\partial}{\partial t}\phi_k^{\bm q}(\bm{r},t)=[\hat{h}^{\bm q}(\bm{r},t)-\varepsilon_k^{\bm q}(t)]\phi_k^{\bm q}(\bm{r},t),
\end{equation}
where $\varepsilon_k^{\bm q}(t)=\langle\psi_k^{\bm q}|\hat{h}^{\bm q}|\psi_k^{\bm q}\rangle$ is the expectation value of the single-particle Hamiltonian $\hat{h}^{\bm q}(\bm{r},t)$, that is self-consistently determined
at each step in time by time-dependent densities and currents in the scalar, vector, and isovector channels,
\begin{subequations}\label{Eq_density_current}
  \begin{align}
    &\rho_S^{\bm q}(\bm{r},t)=\sum_k^{l_q} n_{{\bm q},k}(t)\bar{\phi}^{\bm q}_k(\bm{r},t)\phi^{\bm q}_k(\bm{r},t),\\
    &j^{{\bm q},\mu}(\bm{r},t)=\sum_k^{l_q} n_{{\bm q},k}(t)\bar{\phi}^{\bm q}_k(\bm{r},t)\gamma^\mu\phi^{\bm q}_k(\bm{r},t),\\
    &j_{TV}^{{\bm q},\mu}(\bm{r},t)=\sum_k^{l_q}  n_{{\bm q},k}(t)\bar{\phi}^{\bm q}_k(\bm{r},t)\gamma^\mu\tau_3\phi^{\bm q}_k(\bm{r},t),
  \end{align}
\end{subequations}
where $l_q$ is the number of the canonical basis states, $\tau_3$ is the isospin Pauli matrix.
The time evolution of the occupation probability $n_{{\bm q},k}(t)=|\nu_{{\bm q},k}(t)|^2$, and pairing tensor $\kappa_{{\bm q},k}(t)=\mu_{{\bm q},k}^{*}(t)\nu_{{\bm q},k}(t)$, is governed by the following equations:
\begin{subequations}\label{Eq_td_nkapp_eq_BCS}
   \begin{align}
     &i\frac{d}{dt}n_{{\bm q},k}(t)=\kappa_{{\bm q},k}(t)\Delta_{{\bm q},k}^*(t)-\kappa_{{\bm q},k}^{*}(t)\Delta_{{\bm q},k}(t),\\
     &i\frac{d}{dt}\kappa_{{\bm q},k}(t)=[\varepsilon_k^{\bm q}(t)+\varepsilon_{\bar{k}}^{\bm q}(t)]\kappa_{{\bm q},k}(t)+\Delta_{{\bm q},k}(t)[2n_{{\bm q},k}(t)-1] ,
   \end{align}
\end{subequations}
(for details, see Refs.~\cite{Scamps13TDBCS,ebata10TDBCS}). In time-dependent calculations, a monopole pairing interaction is employed, and the gap parameter $\Delta_{{\bm q},k}(t)$ is defined in terms of single-particle energies and the pairing tensor,
\begin{equation}
  \Delta_{{\bm q},k}(t)=\left[G\sum_{k'>0}f(\varepsilon_{k'}^{\bm q})\kappa_{{\bm q},k'}\right]f(\varepsilon_k^{\bm q}),
\end{equation}
where $f(\varepsilon_k^{\bm q})$ is the cut-off function for the pairing window \cite{Scamps13TDBCS}, and $G$ is the pairing strength.

\subsection{Time evolution of the weight functions $f_{q}(t)$}
The equation of motion for the weight functions is obtained from the time-dependent variational principle~\cite{Regnier2019PRC},
\begin{equation}
    \sum_{\bm q} i\hbar \mathcal{N}_{\bm{q'q}}(t)\partial_t f_{\bm q}(t)=\sum_q \mathcal{H}_{\bm {q'q}}(t)f_{\bm q}(t)-\sum_q \mathcal{H}_{\bm {q'q}}^{MF}(t) f_{\bm q}(t)
    \label{TD-HW-f}
\end{equation}
where the time-dependent kernels
\begin{subequations}
 \begin{align}
&\mathcal{N}_{\bm{q'q}}(t)=\langle\Phi_{\bm {q'}}(t)|\Phi_{\bm q}(t)\rangle,\label{Eq_N}\\
&\mathcal{H}_{\bm{q'q}}(t)=\langle\Phi_{\bm {q'}}(t)|\hat{H}|\Phi_{\bm q}(t)\rangle,\label{Eq_H}\\
&\mathcal{H}^{MF}_{\bm{q'q}}(t)=\langle\Phi_{\bm {q'}}(t)|i\hbar\partial_t|\Phi_{\bm q}(t)\rangle,
\label{Eq_H_mf}
 \end{align}
\end{subequations}
include the overlap, the Hamiltonian, and the time derivative of the generator states, respectively.
The expressions used to calculate the time-dependent kernels were introduced in our previous work~\cite{Li2024FOP} and,
for completeness, are also included in the following sections.

\subsection{Overlap kernel $\mathcal{N}_{\bf {q'q}}(t)$}

According to Eq. (\ref{Eq_Slater}), the expression for the overlap kernel Eq. (\ref{Eq_N}) can be written in the following form:
\begin{equation}\label{Eq_norm_kernel}
    \begin{split}
    \mathcal{N}_{\bm {q'q}}(t)&=\langle\Phi_{\bm {q'}}(t)|\Phi_{\bm q}(t)\rangle\\
    &=\langle-|\prod_{k'>0}[\mu_{{\bm q'},k'}^*(t)+\nu_{{\bm q'},k'}^*(t)c_{{\bm q'},\bar{k}'}(t)c_{{\bm q'},k'}(t)]
    \prod_{k>0}[\mu_{{\bm q},k}(t)+\nu_{{\bm q},k}(t)c_{{\bm q},k}^\dagger(t)c_{{\bm q},\bar{k}}^\dagger(t)]|-\rangle\\
    &=\frac{(-1)^{(l_{\bm q'}-1)l_{\bm q'}/2}}{\prod_{k'}^{l_{\bm q'}/2} \prod_k^{l_{\bm q}/2} \nu_{{\bm q'},k'}^{*}\nu_{{\bm q},k}}\langle-|\beta_{{\bm q'},1}^\dagger...\beta_{{\bm q'},l_{\bm q'}}^\dagger\beta_{{\bm q},1}...\beta_{{\bm q},l_{\bm q}}|-\rangle, \\
    \end{split}
\end{equation}
where $\beta^\dagger_{{\bm q},k}$ is the quasi-particle creation operator associated with the quasiparticle vacuum $|\Phi_{\bf q}(t)\rangle$.
The overlap between two quasi-particle vacua can be calculated using the Pfaffian algorithms
developed in Refs.~\cite{Robledo2009PRC,Hu2014PLB}.

\subsection{Energy kernel $\mathcal{H}_{\bf {q'q}}(t)$}

For the point-coupling relativistic energy density functional PC-PK1 \cite{Zhao2010PRC},
one obtains the expression for the energy kernel $\mathcal{H}^{\rm DF}(t)$, under the assumption \cite{nakatsukasa16} that it only depends on the transition densities at time $t$:
\begin{equation}
\begin{aligned}
        \mathcal{H}^{{\rm DF}}_{\bm {q'q}}(t)=\langle\Phi_{\bm {q'}}(t)|\hat{H}^{{\rm DF}}|\Phi_{\bm q}(t)\rangle&=\langle\Phi_{\bm {q'}}(t)|\Phi_{\bm q}(t)\rangle\cdot\int d^3r~\{ \rho_{\rm kin}(\bm{r},t)\\
        &+\frac{\alpha_S}{2}\rho_S(\bm{r},t)^2+\frac{\beta_S}{3}\rho_S(\bm{r},t)^3\\
        &+\frac{\gamma_S}{4}\rho_S(\bm{r},t)^4+\frac{\delta_S}{2}\rho_S(\bm{r},t)\Delta\rho_S(\bm{r},t)\\
        &+\frac{\alpha_V}{2} j^\mu(\bm{r},t) j_\mu(\bm{r},t)+\frac{\gamma_V}{4}(j^\mu(\bm{r},t) j_\mu(\bm{r},t))^2\\
        &+\frac{\delta_V}{2} j^\mu(\bm{r},t)\Delta j_\mu(\bm{r},t)+\frac{\alpha_{TV}}{2} j^\mu_{TV}(\bm{r},t)\cdot [j_{TV}(\bm{r},t)]_\mu\\
        &+\frac{ \delta_{TV}}{2} j^\mu_{TV}(\bm{r},t)\cdot \Delta[j_{TV}(\bm{r},t)]_\mu+\frac{e^2}{2}j^\mu_p(\bm{r},t) A_\mu(\bm{r},t) \},
\end{aligned}
\end{equation}
where the densities and currents~$\rho_{\rm kin}$,~$\rho_S$,~$j^\mu$,~$j_{TV}^\mu$,~and~$j_p^\mu$ read
\begin{subequations}
 \begin{align}
        &\rho_{\rm kin}(\bm{r},t)=\sum_{k'}^{l_{\bm q'}}\sum_{k}^{l_{\bm q}}\bar{\phi}_{k'}^{\bm {q'}}(\bm{r},t)(-i\bm{\gamma}\cdot\bm{\nabla}+m_N)\phi_{k}^{\bm q}(\bm{r},t)\rho^{\rm tran}_{{\bm {q'q}},k'k}(t),\\
        &\rho_S(\bm{r},t)=\sum_{k'}^{l_{\bm q'}}\sum_{k}^{l_{\bm q}}\bar{\phi}_{k'}^{\bm {q'}}(\bm{r},t)\phi_{k}^{\bm q}(\bm{r},t)\rho^{\rm tran}_{{\bm {q'q}},k'k}(t),\\
        &j^\mu(\bm{r},t)=\sum_{k'}^{l_{\bm q'}}\sum_{k}^{l_{\bm q}}\bar{\phi}_{k'}^{\bm {q'}}(\bm{r},t)\gamma^\mu\phi_{k}^{\bm q} (\bm{r},t)\rho^{\rm tran}_{{\bm {q'q}},k'k}(t),\\
        &j_{TV}^\mu(\bm{r},t)=\sum_{k'}^{l_{\bm q'}}\sum_{k}^{l_{\bm q}}\bar{\phi}_{k'}^{{\bm q'}}(\bm{r},t)\tau_3\gamma^\mu\phi_{k}^{\bm q}(\bm{r},t)\rho^{\rm tran}_{{\bm {q'q}},k'k}(t),\\
        &j_p^\mu(\bm{r},t)=\frac{1-\tau_3}{2}\sum_{k'}^{l_{\bm q'}}\sum_{k}^{l_{\bm q}}\bar{\phi}_{k'}^{\bm {q'}}(\bm{r},t)\gamma^\mu\phi_{k}^{\bm q}(\bm{r},t)\rho^{\rm tran}_{{\bm {q'q}},k'k}(t).
\end{align}
\end{subequations}
\bigskip

The transition density matrix $\rho^{\rm tran}(t)$ is defined by the following relation
\begin{equation}
\rho^{\rm tran}_{{\bm {q'q}},k'k}(t)=\frac{ \langle\Phi_{\bm {q'}}(t)|c_{{\bm {q'}},k'}^{\dagger}(t)c_{{\bm q},k}(t)|\Phi_{\bm q}(t)\rangle}{\langle\Phi_{\bm {q'}}(t)|\Phi_{\bm q}(t)\rangle}
=\nu_{{\bm q'},k'}^{*}(t)\nu_{{\bm q},k}(t)\frac{ \langle\Phi_{\bm {q'}}(t)|\beta_{{\bm {q'}},\bar{k}'}(t)\beta_{{\bm q},\bar{k}}^{\dagger}(t)|\Phi_{\bm q}(t)\rangle}{\langle\Phi_{\bm {q'}}(t)|\Phi_{\bm q}(t)\rangle}.
\end{equation}
The numerator of the transition density matrix $\rho^{\rm tran}_{{\bm {q'q}},k'k}(t)$ is the overlap between a quasi-particle vacuum with $\left(l_{\bm q'}-1\right)$ quasi-particle levels and a quasi-particle vacuum with $\left(l_{\bm q}-1\right)$ quasi-particle levels.
It can be calculated using the Pfaffian algorithms~\cite{Hu2014PLB,Robledo2009PRC}.

For monopole pairing, the pairing Hamiltonian operator $\hat{H}^{{\rm pair}}$ in 3D-lattice space is defined:
\begin{equation}
\hat{H}_{pair}=-\sum_{r_1,s_1>0,r_2,s_2>0} G(c^\dag_{r_1,s_1}c^\dag_{r_1,\bar{s}_1})(c_{r_2,\bar{s}_2}c_{r_2,{s}_2})
\end{equation}
where $c^\dag_{r_1,s_1}$ is the creation operator for the lattice coordinate wave function $|r_1,s_1\rangle$, and $r_1$ is the index of the lattice point, and $s_1$ is the index of the spin.
One obtains the expression for the pairing part of the energy kernel $\mathcal{H}^{\rm pair}(t)$ in 3D-lattice space
\begin{equation}
\begin{aligned}
        &\mathcal{H}^{{\rm pair}}_{\bm {q'q}}(t)=\langle\Phi_{\bm {q'}}(t)|\hat{H}^{{\rm pair}}|\Phi_{\bm q}(t)\rangle\\
        &=- G ~\langle\Phi_{\bm {q'}}(t)|\Phi_{\bm q}(t)\rangle \sum_{k_1,k_2,k_3,k_4>0}[f(\varepsilon_{k_1}^{\bm q'})f(\varepsilon_{k_2}^{\bm q})f(\varepsilon_{k_3}^{\bm q'})f(\varepsilon_{k_4}^{\bm q})]^{1/2}  \\
        &\times \langle\phi^{\bm q'}_{k_1}|\phi^{\bm q}_{k_2}\rangle\langle\phi^{\bm q'}_{k_3}|\phi^{\bm q}_{k_4}\rangle
         [\kappa_{\bm{q'q},\bar{k}_2k_1}^{\rm tran}(t)]^{*}\kappa_{\bm{qq'},\bar{k}_3k_4}^{\rm tran}(t),\\
\end{aligned}
\end{equation}
where the transition pairing tensor matrix $\kappa^{\rm tran}(t)$ is defined by the following relation
\begin{equation}
\kappa^{\rm tran}_{{\bm {q'q}},k'k}(t)=\frac{ \langle\Phi_{\bm {q'}}(t)|c_{{\bm {q'}},k'}(t)c_{{\bm q},k}(t)|\Phi_{\bm q}(t)\rangle}{\langle\Phi_{\bm {q'}}(t)|\Phi_{\bm q}(t)\rangle}
=\mu_{{\bm q'},k'}^{*}(t)\nu_{{\bm q},k}(t)\frac{ \langle\Phi_{\bm {q'}}(t)|\beta_{{\bm {q'}},k'}(t)\beta_{{\bm q},\bar{k}}^{\dagger}(t)|\Phi_{\bm q}(t)\rangle}{\langle\Phi_{\bm {q'}}(t)|\Phi_{\bm q}(t)\rangle}.
\end{equation}

In the BCS model, $|\Phi_{\bm q}(t)\rangle$ is not an eigenstate of the neutron (proton) number operator $\hat{N}$ ($\hat{Z}$), and its expectation value in the collective wave function generally deviates from the desired neutron number $N_0$ (proton number $Z_0$). The method developed in Ref.~\cite{Bonche1990NPA} is used to correct for variations of the nucleon number. The energy kernel finally reads
\begin{equation}
\mathcal{H}_{\bm {q'q}}(t)= \mathcal{H}^{{\rm DF}}_{\bm {q'q}}(t)+ \mathcal{H}^{{\rm pair}}_{\bm {q'q}}(t)
-\lambda_{N}^{\bm {q'q}}(t)[\langle\Phi_{\bm {q'}}(t)|\hat{N}|\Phi_{\bm q}(t)\rangle-N_0]-\lambda_{Z}^{\bm {q'q}}(t)[\langle\Phi_{\bm {q'}}(t)|\hat{Z}|\Phi_{\bm q}(t)\rangle-Z_0],
\end{equation}
where $\lambda_{i}^{\bm {q'q}}(t)$ is defined as the average of the chemical potentials $\lambda_{i}^{\bm {q'}}(t)$ and $\lambda_{i}^{\bm {q}}(t)$,
\begin{equation}
\lambda_{i}^{\bm {q'q}}(t)=\frac{\lambda_{i}^{\bm {q'}}(t)+\lambda_{i}^{\bm {q}}(t)}{2},~~~i=N,Z.
\end{equation}

\subsection{Mean-field kernel $\mathcal{H}^{MF}_{\bf {q'q}}(t)$}
From the expression for the time evolution of $|\Phi_{\bm q}(t)\rangle$~\cite{Ren_22PRL,Ren_22PRC},
\begin{equation}\label{MF_kernel}
 \begin{aligned}
    &i\hbar\partial_t|\Phi_{\bm q}(t)\rangle\\
    &=i\hbar\sum_{k>0}\{\partial_t[c_{{\bm q},k}^{\dagger}(t)c_{{\bm q},\bar{k}}^\dagger(t)]+\dot{\mu}_{{\bm q},k}(t)+\dot{\nu}_{{\bm q},k}(t)c_{{\bm q},k}^\dagger(t)c_{{\bm q},\bar{k}}^\dagger(t)\}
    \prod_{j\neq k,j>0}[\mu_{{\bm q},j}(t)+\nu_{{\bm q},j}(t)c_{{\bm q},j}^\dagger(t)c_{{\bm q},\bar{j}}^\dagger(t)]|-\rangle\\
    &=\sum_{k}^{l_q}[\hat{h}^{\bm q}(\bm{r},t)-\varepsilon_k^{\bm q}(t)]c_{{\bm q},k}^{\dagger}(t)c_{{\bm q},k}(t)|\Phi_{\bm q}(t)\rangle
   +i\hbar\sum_{k>0}\sqrt{|\dot{\mu}_{{\bm q},k}(t)|^2+|\dot{\nu}_{{\bm q},k}(t)|^2}~|\tilde{\Phi}_{{\bm q},k}(t)\rangle,
 \end{aligned}
\end{equation}
where the Slater determinant $|\tilde{\Phi}_{{\bm q},k}(t)\rangle$ is defined as
\begin{equation}
 \begin{aligned}
|\tilde{\Phi}_{{\bm q},k}(t)\rangle&=[ \frac{\dot{\mu}_{{\bm q},k}(t)}{\sqrt{|\dot{\mu}_{{\bm q},k}(t)|^2+|\dot{\nu}_{{\bm q},k}(t)|^2}}
                        +\frac{\dot{\nu}_{{\bm q},k}(t)}{\sqrt{|\dot{\mu}_{{\bm q},k}(t)|^2+|\dot{\nu}_{{\bm q},k}(t)|^2}}c_{{\bm q},k}^\dagger(t)c_{{\bm q},\bar{k}}^\dagger(t)]\\
                        &\cdot\prod_{j\neq k,j>0}[\mu_{{\bm q},j}(t)+\nu_{{\bm q},j}(t)c_{{\bm q},j}^\dagger(t)c_{{\bm q},\bar{j}}^\dagger(t)]|-\rangle
\end{aligned}
\end{equation}
Eq. (\ref{Eq_H_mf}) can be written in the form
\begin{equation}
 \begin{aligned}
\mathcal{H}^{MF}_{\bm {q'q}}(t)&=\langle\Phi_{\bm q'}(t)|i\hbar\partial_t|\Phi_{\bm q}(t)\rangle\\
&=\langle\Phi_{\bm q'}(t)|\sum_{k}^{l_q}[\hat{h}^{\bm q}(\bm{r},t)-\varepsilon_k^{\bm q}(t)]c_{{\bm q},{k}}^{\dagger}(t)c_{{\bm q},{k}}(t)|\Phi_{\bm q}(t)\rangle\\
&+i\hbar\sum_{k>0}\sqrt{|\dot{\mu}_{{\bm q},k}(t)|^2+|\dot{\nu}_{{\bm q},k}(t)|^2}~\langle\Phi_{\bm q'}(t)|\tilde{\Phi}_{{\bm q},k}(t)\rangle.
\end{aligned}
\end{equation}
By expanding $[\hat{h}^{\bm q}(\bm{r},t)-\varepsilon_k^{\bm q}(t)]c^{\dagger}_{{\bm q},k}(t)$ in a complete basis $ c^{\dagger}_{{\bm q'},k'}(t)$,
\begin{equation}
     [\hat{h}^{\bm q}(\bm{r},t)-\varepsilon_k^{\bm q}(t)]c^{\dagger}_{{\bm q},k}(t)=\sum_{k'}\langle \phi^{{\bm q'}}_{k'}(\bm{r},t)|[\hat{h}^{\bm q}(\bm{r},t)-\varepsilon_k^{\bm q}(t)]|\phi^{\bm q}_{k}(\bm{r},t)\rangle c^{\dagger}_{{\bm q'},k'}(t),
\end{equation}
one obtains for $\mathcal{H}_{\bm {q'q}}^{MF}(t)$ the expression
\begin{equation}
 \begin{aligned}
\mathcal{H}_{\bm{q'q}}^{MF}(t)&=\langle\Phi_{\bm q'}(t)|\Phi_{\bm q}(t)\rangle\cdot\sum_{k'}^{l_{\bm q'}}\sum_{k}^{l_{\bm q}}\langle \phi^{\bm q'}_{k'}(\bm{r},t)|[\hat{h}^{\bm q}(\bm{r},t)-\varepsilon_k^{\bm q}(t)]|\phi^{\bm q}_{k}(\bm{r},t)\rangle\rho^{\rm tran}_{k'k}(t)\\
                              &+i\hbar\sum_{k>0}\sqrt{|\dot{\mu}_{{\bm q},k}(t)|^2+|\dot{\nu}_{{\bm q},k}(t)|^2}~\langle\Phi_{\bm q'}(t)|\tilde{\Phi}_{{\bm q},k}(t)\rangle,
\end{aligned}
\end{equation}
where $\dot{\mu}_{{\bm q},k}(t)$ and $\dot{\nu}_{{\bm q},k}(t)$ can be derived from Eq.~(\ref{Eq_td_nkapp_eq_BCS}),
and $\langle\Phi_{\bm q'}(t)|\tilde{\Phi}_{{\bm q},k}(t)\rangle$  can be obtained by the Pfaffian algorithms~\cite{Hu2014PLB,Robledo2009PRC}.

\subsection{Collective wave function $g(t)$}
The weight function $f_{\bm q}$ is not a probability amplitude of finding the system at the collective coordinate ${\bm q}$,
The corresponding collective wave function $g_{\bm q}(t)$ is defined by the transformation~\cite{Reinhard1987RPP}
\begin{equation}
    g=\mathcal{N}^{1/2}f,
    \label{Eq_f_g}
\end{equation}
where $\mathcal{N}^{1/2}$ is the square root of the overlap kernel matrix. Inserting Eq.~(\ref{Eq_f_g}) into Eq.~(\ref{TD-HW-f}), the time evolution of the collective wave function is governed by the following equation \cite{Regnier2019PRC}
\begin{equation}
\label{Eq_HW_4}
    i\hbar \dot{g}=\mathcal{N}^{-1/2}(H-H^{MF})\mathcal{N}^{-1/2}g+i\hbar\dot{\mathcal{N}}^{1/2}\mathcal{N}^{-1/2}g.
\end{equation}

\subsection{Observables $\hat{O}$}
The kernel of any observable $\hat{O}$
\begin{equation}
 \mathcal{O}_{\bm {q'q}}=\langle\Phi_{q'}(t)|\hat{O}|\Phi_q(t)\rangle
\end{equation}
can be mapped to the corresponding collective operator $\mathcal{O}^c$:
\begin{equation}
\mathcal{O}^c=\mathcal{N}^{-1/2}\mathcal{O}\mathcal{N}^{-1/2}.
\end{equation}
The expectation value of an observable $\hat{O}$ in the correlated nuclear wave function reads
\begin{equation}
\label{Eq_observable}
\langle \Psi(t)|\hat{O}|\Psi(t)\rangle =f^\dag \mathcal{O} f=g^\dag \mathcal{O}^c g.
\end{equation}

\section{SUPPLEMENTAL METHODS: Particle number projection method}

The nuclear wave function is a superposition of a number of Slater determinants (here we omit the $t$-index),
\begin{equation}
    |\Psi\rangle=\sum_{\bm q} f_{ \bm q} |\Phi_{ \bm q}\rangle.
\end{equation}
where $|\Phi_{ \bm q}\rangle$ is the BCS vacuum,
\begin{equation}
\left|\Phi_{ \bm q}\right\rangle=\prod_{k>0}[u_{k,{ \bm q}}+v_{k,{ \bm q}}a_{k,{ \bm q}}^\dagger a_{\bar{k},{ \bm q}}^\dagger]|-\rangle.
\end{equation}
$u_{k,{ \bm q}}$ and $v_{k,{ \bm q}}$ are Bogoliubov transformation coefficients, and $a_{k,{ \bm q}}^\dagger$ is the single particle creation operator.

The probability of finding $n$ particles in the subspace $V_f$, for a nucleus with total particle number $N$ is
\begin{equation}
   P(n|N)=\frac{\langle\Psi\left|\hat{P}_n^{V_f}\hat{P}_N\right|\Psi \rangle}{\langle\Psi \left|\widehat{P}_N\right|\Psi \rangle}=
   \frac{\sum_{\bm q \bm q\prime}f_{{ \bm q}}^{*}f_{{ \bm q}} \langle\Phi_{{ \bm q}\prime}\left|\hat{P}_n^{V_f}\hat{P}_N\right|\Phi_{{ \bm q}}\rangle}%
   {\sum_{\bm q \bm q\prime}f_{{ \bm q}\prime}^{*} f_{{ \bm q}} \langle\Phi_{{ \bm q}\prime} \left|\hat{P}_N \right|\Phi_{{ \bm q}}\rangle}.%
\end{equation}
$\hat{P}_n^{V_f}$ and $\hat{P}_N$ are the projection operators defined in the subspace $V_f$ and the entire space, respectively,
\begin{equation}
    \hat{P}_n^{V_f}=\frac{1}{2\pi}\int_0^{2\pi} d\theta~e^{i\theta(n-\hat{N}_{V_f})},
    \label{P_n_2}
\end{equation}
\begin{equation}
    \hat{P}_N=\frac{1}{2\pi}\int_0^{2\pi} d\theta~e^{i\theta(N-\hat{N})},
    \label{P_N}
\end{equation}
where $\hat{N}_{V_f}$ is the particle number operator in the subspace $V_f$, and $\hat{N}$~is the particle number operator in the entire space.

The $N$-particle number projected state at a collective coordinate ${ \bm q}$ reads
\begin{equation}
    \left|\Phi_{{ \bm q}}^{N}\right\rangle=\widehat{P}_{N}\left|\Phi_{{ \bm q}}\right\rangle=\frac{1}{2\pi}\int_{0}^{2\pi}d\theta~e^{i(\widehat{N}-N)}|\Phi_{{ \bm q}}\rangle,
\end{equation}
and can be expressed in the form given of a contour integral
\begin{equation}
    \left|\Phi_{{ \bm q}}^{N}\right\rangle=\frac{1}{2\pi i}\oint_{C} dz~z^{\widehat{N}-N-1}\big|\Phi_{{ \bm q}}\big\rangle,
    \label{Eq_Phi_N_z}
\end{equation}
where $C$ is an arbitrary closed contour encircling the origin $z=0$ of the complex plane.
As shown in Ref.~\cite{Dobaczewski2007PRC}, one can defined a shift operator
\begin{equation}
    \hat{z}(z)=z^{\hat{N}}=e^{(\eta+i\theta)\hat{N}},
\end{equation}
parametrized by means of a single complex number $z$, $\ln(z)=\eta+i\theta$.
The shift operator constitutes a non-unitary transformation,
\begin{equation}
    \hat{z}a_{k,{\bm q}}^{\dagger}\hat{z}^{-1}=za_{k,{\bm q}}^{\dagger},\quad\hat{z}a_{k,{\bm q}}\hat{z}^{-1}=z^{-1}a_{k,{\bm q}}.
\end{equation}
Obviously, for $z=1$, the shift operator is equal to identity.

The kernel $\langle\Phi_{{ \bm q}^{\prime}}\big|\hat{P}_{N}\big|\Phi_{q}\rangle $ can be evaluated from
\begin{equation}
    \langle\Phi_{{ \bm q}^{\prime}}\big|\hat{P}_{N}\big|\Phi_{{ \bm q}}\rangle= \langle\Phi_{{ \bm q}^{\prime}}\big|\Phi_{{ \bm q}}^{N}\rangle=\frac{1}{2\pi i}\oint_{C}dz~z^{-N-1}\langle\Phi_{{ \bm q}^{\prime}}\big|\Phi_{{ \bm q}}(z)\big\rangle,
    \label{Eq_P_N_z}
\end{equation}
where shifted state at a collective coordinate ${ \bm q}$ takes the form
\begin{equation}
    \begin{aligned}
    &\left|\Phi_{{ \bm q}}(z)\right\rangle=\hat{z}\big|\Phi_{{ \bm q}}\big\rangle=\hat{z}\prod_{k>0}(u_{k,{ \bm q}}+v_{k,q}a_{k,{ \bm q}}^{\dagger}a_{\bar{k},{ \bm q}}^{\dagger})|-\rangle\\
    &=\prod_{k>0}(u_{k,q}+v_{k,{ \bm q}}\hat{z}a_{k,{ \bm q}}^{\dagger}\hat{z}^{-1}\hat{z}a_{\bar{k},{ \bm q}}^{\dagger}\hat{z}^{-1})\hat{z}|-\rangle
    =\prod_{k>0}(u_{k,{ \bm q}}+v_{k,{ \bm q}}z^{2}a_{k,{ \bm q}}^{\dagger}a_{\bar{k},{ \bm q}}^{\dagger})|-\rangle.
    \end{aligned}
\end{equation}
When the closed contour $C$ is chosen as a unit circle $z=e^{i\theta}$, Eq.~(\ref{Eq_P_N_z}) is equivalent to
\begin{equation}
    \langle\Phi_{{ \bm q}^{\prime}}\big|\hat{P}_{N}\big|\Phi_{{ \bm q}}\rangle=\frac{1}{2\pi } \int_0^{2\pi} d\theta~e^{-iN\theta}\langle\Phi_{{ \bm q}^{\prime}}\big|\Phi_{{ \bm q}}(\theta)\big\rangle,
    \label{Eq_P_N}
\end{equation}
where $|\Phi_{{ \bm q}}(\theta)\big\rangle$ is a BCS vacuum with the Bogoliubov transformation coefficients $u_{k,{\bm q}}$ and $e^{2i\theta} v_{k,{\bm q}}$
and the overlap $\langle\Phi_{{\bm q}^{\prime}}\big|\Phi_{{\bm q}}(\theta)\big\rangle$ can be calculated by the Pfaffian algorithms~\cite{Hu2014PLB,Robledo2009PRC}.

The particle number projected state at a collective coordinate ${\bm q}$, which corresponds to $n$ particles in the subspace $V_f$ and $N$ particles in the full space, is defined as
\begin{equation}
    \left|\Phi_{{\bm q}}^{N,n}\right\rangle=\hat{P}_{n}^{V_f}\hat{P}_{N}\big|\Phi_{{\bm q}}\big\rangle=\hat{P}_{n}^{V_f}\big|\Phi_{{\bm q}}^{N}\big\rangle
    =\frac{1}{2\pi}\int_{0}^{2\pi}d\theta~e^{i(\widehat{N}_{V_f}-n)}\big|\Phi_{{\bm q}}^{N}\big\rangle.
\end{equation}
A more general form given by the contour integral reads
\begin{equation}
    \big|\Phi_{{\bm q}}^{N,n}\big\rangle=\frac{1}{2\pi i}\oint_{C^{\prime}}dz^{\prime}~z^{\prime\hat{N}_{V_f}-n-1}\big|\Phi_{{\bm q}}^{N}\big\rangle,
\end{equation}
where $C'$ is an arbitrary closed contour encircling the origin $z^{\prime}=0$ of the complex plane.
The shift operator is defined by
\begin{equation}
    \hat{z}^{\prime}(z^{\prime})=z^{\prime\hat{N}_{V_f}}=e^{\left(\eta^{\prime}+i\theta^{\prime}\right)\hat{N}_{V_f}},
\end{equation}
and is parametrized by means of a single complex number $z^{\prime}$, $\ln(z^{\prime})=\eta^{\prime}+i\theta^{\prime}$.
To simplify the calculations, a set of operators is defined,
\begin{equation}
    b_{k,q}^{\dagger} (z^{\prime})=\hat{z}^{\prime}a_{k,{\bm q}}^{\dagger}\hat{z}^{\prime-1},\quad b_{k} (z^{\prime})=(\hat{z}^{\prime-1})^{\dagger}a_{k,{\bm q}}\hat{z}^{\prime\dagger}.
    \label{Eq_b_k}
\end{equation}

The kernel $\langle\Phi_{{\bm q}^{\prime}}\left|\widehat{P}_{n}^{V_{f}}\hat{P}_{N}\right|\Phi_{{\bm q}}\rangle$ can be evaluated using the expression
\begin{equation}
    \langle\Phi_{{\bm q}^{\prime}}\left|\hat{P}_{n}^{V_{f}}\hat{P}_{N}\right|\Phi_{{\bm q}}\rangle=\langle\Phi_{{\bm q}^{\prime}} |\Phi_{{\bm q}}^{N,n} \rangle
    =\frac{1}{2\pi i}\oint_{C}dz~z^{-N-1}\frac{1}{2\pi i}\oint_{C^{\prime}}dz^{\prime}~z^{\prime-n-1}\langle\Phi_{{\bm q}^{\prime}}\big|\Phi_{{\bm q}}(z,z^{\prime})\big\rangle,
    \label{Eq_P_Nn_z}
\end{equation}
where shifted state $\big|\Phi_{{\bm q}}(z,z^{\prime})\big\rangle$ reads
\begin{equation}
    \begin{aligned}
    &\left|\Phi_{{\bm q}}(z^{\prime},z)\right\rangle=\hat{z}^{\prime}\hat{z}\big|\Phi_{{\bm q}}\big\rangle=\hat{z}^{\prime}\prod_{k>0}(u_{k,{\bm q}}+z^{2}v_{k,{\bm q}}a_{k,{\bm q}}^{\dagger}a_{\bar{k},{\bm q}}^{\dagger})|-\rangle\\
    &=\prod_{k>0}(u_{k,{\bm q}}+v_{k,{\bm q}}z^{2}\hat{z}^{\prime}a_{k,{\bm q}}^{\dagger}\hat{z}^{\prime-1}\hat{z}^{\prime}a_{\bar{k},{\bm q}}^{\dagger}\hat{z}^{\prime-1})\hat{z}^{\prime}|-\rangle
    =\prod_{k>0}(u_{k,q}+v_{k,{\bm q}}z^{2}b_{k,{\bm q}}^{\dagger}(z^{\prime})b_{\bar{k},{\bm q}}^{\dagger}(z^{\prime}))|-\rangle.
    \end{aligned}
\end{equation}
When both closed contours, $C$ and $C^{\prime}$, are chosen as unit circles $z=e^{i\theta}$ and $z^{\prime}=e^{i\theta^{\prime}}$,  Eq.~(\ref{Eq_P_Nn_z}) is equivalent to
\begin{equation}
 \langle\Phi_{{\bm q}^{\prime}}\left|\hat{P}_{n}^{V_{f}}\hat{P}_{N}\right|\Phi_{{\bm q}}\rangle=\frac{1}{2\pi } \int_0^{2\pi} d\theta~e^{-iN\theta} \frac{1}{2\pi} \int_0^{2\pi} d\theta^{\prime}~e^{-in\theta^{\prime}}
 \langle\Phi_{{\bm q}^{\prime}}\big|\Phi_{{\bm q}}(\theta,\theta^{\prime})\big\rangle,
 \label{Eq_P_Nn}
\end{equation}
where $\left|\Phi_{\bm q}(\theta,\theta^{\prime})\right\rangle$ reads
\begin{equation}
    \left|\Phi_{{\bm q}}(\theta,\theta^{\prime})\right\rangle=\prod_{k>0}(u_{k,q}+v_{k,{\bm q}}e^{2i\theta}b_{k,{\bm q}}^{\dagger}(e^{i\theta^{\prime}})b_{\bar{k},{\bm q}}^{\dagger}(e^{i\theta^{\prime}}))|-\rangle.
\end{equation}
The operators $b_{k,{\bm q}}^{\dagger}(\theta^{\prime})$ and $b_{k,{\bm q}}(\theta^{\prime})$ are obtained by Eq.~(\ref{Eq_b_k}),
\begin{equation}
    b_{k,{\bm q}}^{\dagger}(e^{i\theta^{\prime}})=e^{i\theta^{\prime}\hat{N}_{V_f}}a_{k,{\bm q}}^{\dagger}e^{-i\theta^{\prime}\hat{N}_{V_f}},
     b_{k,{\bm q}}(e^{i\theta^{\prime}})=e^{i\theta^{\prime}\hat{N}_{V_f}}a_{k,{\bm q}}e^{-i\theta^{\prime}\hat{N}_{V_f}},
\end{equation}
and satisfy the orthonormality condition,
\begin{equation}
\langle-|b_{k,{\bm q}}(\theta^{\prime}) b_{k',{\bm q}}^{\dagger}(\theta^{\prime})|-\rangle=\langle-|a_{k,{\bm q}}a_{k',{\bm q}}^{\dagger}|-\rangle=\delta_{kk'}.
\end{equation}
So, $\left|\Phi_{\bm q}(\theta,\theta^{\prime})\right\rangle$ is equivalent to a BCS vacuum, and overlap $\langle\Phi_{{\bm q}^{\prime}}|\Phi_{{\bm q}}(\theta,\theta^{\prime})\rangle$ can be calculated using the Pfaffian algorithms~\cite{Hu2014PLB,Robledo2009PRC}.

In the present calculations that employ the generalized TD-GCM, the mesh spacing of the lattice is 1.0 fm for all directions, and the box size is $L_x\times L_y\times L_z=20\times20\times60~{\rm fm}^3$.
The step for the time evolution is $0.2~{\rm fm}/c~\approx 6.67\times10^{-4}$~zs.
The energy surface and initial states for the time evolution are obtained by self-consistent deformation-constrained relativistic DFT calculations in a three-dimensional lattice space~\cite{Ren2017PRC,Ren2017PRC, ren19LCS, ren20_NPA,Li2020PRC,Xu2024PRC,Xu2024PLB} with the box size $L_x\times L_y\times L_z=20\times20\times50~{\rm fm}^3$.
Both static and dynamical calculations are based on the relativistic density functional PC-PK1~\cite{Zhao2010PRC}, together with a monopole pairing interaction.
Pairing correlations are taken into account in the Bardeen-Cooper-Schrieffer (BCS) approximation.
The pairing strength parameters: $-0.135$ MeV for neutrons, and $-0.230$ MeV for protons, are determined by the empirical pairing gaps of $^{240}$Pu, using the three-point odd-even mass formula
The subspace $V_f$ is selected as the $z>0$ space, where the $z$-axis is along the fission direction, and the interval of $z$ is from $-30$~fm to $30$~fm.
In the calculation of~Eqs.~(\ref{Eq_P_N}) and (\ref{Eq_P_Nn}), we evaluate the integrals over $\theta$ and $\theta^{\prime}$ by employing the trapezoidal rule, discretizing the interval into 200 grids.

\bigskip

%

\end{document}